# Reaction barriers and deformation energies of $C_{60}$-based composites

**Elena F Sheka and Landysh Kh Shaymardanova**
Peoples' Friendship University of Russia
Miklukho-Maklaya 6, Moscow 117198, Russia

E-mail: sheka@icp.ac.ru

**Abstract.** The current paper is aimed at the determination of barriers that govern the covalent coupling between two fullerenes $C_{60}$ ($C_{60}$ dimer), $C_{60}$ and single-walled carbon nanotube ([$C_{60}$-(4,4)] carbon nanobud), and $C_{60}$ and graphene ([$C_{60}$-(5,5)] and [$C_{60}$-(9,8)] graphene nanobuds). *Brutto* barriers determined as couplings energies $E_{cpl}^{tot}$ are expanded over two contributions that present total energy of deformation of the composites' components $E_{def}^{tot}$ and energy of covalent coupling $E_{cov}^{tot}$. In view of these energetic parameters and in contrast to expectations, seemingly identical reactions result in different final products. The peculiarity is suggested to be provided by a topochemical character of the covalent coupling between any two members of the $sp^2$ nanocarbons' family. The computations were performed by using the AM1 semiempirical version of unrestricted broken symmetry Hartree-Fock approach.

**Introduction**
Recently started manufacturing of $sp^2$ nanocarbon-based composite materials pursues well-defined goals to provide the best conditions for the exhibition and practical utilization of extraordinary thermal, mechanic, electronic, and chemical properties of the nanocarbons. Obvious success in achieving the goal in the case of fullerenes, carbon nanotubes (CNTs), and graphene dissolved in different polymers points to great perspectives of a new class of composites and their use in a variety of applications (see [1] and references therein). Since low-concentration solutions of individual fullerenes, CNTs, and graphene flakes can be obtained, one can put a question what can we expect when both fullerene and CNTs or graphene are dissolved simultaneously? First experimental attempts to reach the goal have been successful. A few techniques have been suggested to obtain $C_{60}$+CNT composites in which fullerene is located either inside (see review [2] and references therein) or outside [3-6] the CNT wall. Terms peapod [2] and 'nanobud' [4] were suggested to distinguish the two configurations. However, until recently there has been no information concerning the creation of $C_{60}$+graphene composites while those related to CNT+graphene ones seem to become quite known [7,

8]. Is this connected with the difference in chemical reactions in the two cases or are there some other reasons for complication of the composite formation in the former case? In the current paper we shall try to answer the question from the position of energetic parameters that govern addition reactions between fullerene $C_{60}$ as the main partner and $C_{60}$, single-wall carbon nanotube, and nanographenes presenting the second partner of the studies composites.

From the basic standpoint the question addresses the problem of the intermolecular interaction (IMI) between the components. In all the cases, the IMI is greatly contributed with the donor-acceptor (DA) interaction since all the above $sp^2$ nanocarbons are simultaneously good donors and acceptors of electron [9-11]. Within the framework of general characteristics of the DA interaction, the IMI term configuration in the ground state depends on the difference of the asymptotes of $E_{int}(A^0B^0)$ and $E_{int}(A^+B^-)$ terms that describe interaction between neutral molecule and molecular ions, $E_{gap} = I_A - \varepsilon_B$. Here $I_A$ and $\varepsilon_B$ present ionization potential and electron affinity of components $A$ and $B$, respectively. When $E_{gap}$ is as big as it in the case of $C_{60}$ dimers, the IMI term of the ground state has a typical two-well shape shown in figure 1 [10]. The formation of a stationary product $AB$ at point $R^{(+-)}$ is accompanied by the creation of 'intermolecular' chemical bonds between $A$ and $B$ partners. Oppositely, widely spaced neutral moieties form a charge transfer complex $A+B$ in the vicinity of point $R^{(00)}$ [9-11]. In spite of the formation of $AB$ product is energetically profitable, the yield of the relevant reaction when starting from $A+B$ mixture is determined by a barrier that separates $A+B$ and $AB$ products. The current paper is aimed at the barrier determination in the case of three types of composites, namely, $C_{60}+C_{60}$, $C_{60}+CNT$, and $C_{60}+$graphene. Since both ionization potentials and electron affinities of $C_{60}$, CNT, and graphene are quite similar by value, the above composites are characterized by practically the same $E_{gap}$. Beside $E_{gap}$, the discussed composites are characterized by alike configuration of the contact place. The two factors make it possible to expect a similar behavior of the addition reactions between the composites partner followed by similar profiles of the barrier energy. As turned out, this expectation was not proven by the calculations performed in the current study, thus highlighting that not only $E_{gap}$ but other factors influence the barrier profiles, which makes practical reactions for the three types of composites quite different. The computations were carried out by using the AM1 semiempirical version of unrestricted broken symmetry Hartree-Fock approach.

## 2. Ground-state term of the $C_{60}$-$C_{60}$ dyad

According to the scheme in figure 1, the reaction of the $C_{60}$ dimer formation can be considered as moving the two molecules towards each other, once spaced initially at large intermolecular distance $R$, then equilibrated and slightly coupled as $A+B$ complex in the $R^{00}$ minimum and afterwards achieved minimum at $R^{+-}$ to form covalently bound adduct $AB$. The last stage implies overcoming a barrier, which is followed by the transition from $(A^0B^0)$ to $(A^+B^-)$ branch of terms after which Coulomb interaction between molecular ions completes the formation of the final $AB$ adduct at the $R^{+-}$ minimum. As shown [9], neither ionization nor positive charging of $C_{60}$ causes lengthening of the molecule valence bonds more than by 0.02 Å. The same can be attributed to CNTs and graphene. Therefore, the ion formation is not accompanied by a noticeable shift of the atom equilibrium positions along any internal coordinate and does not cause any significant vibrational excitation during the relevant transition that could have caused the molecule decomposition under ionization. That is why the fullerene dimerization and/or oligomerization as well as the formation of both CNBs and GNBs should occur as a direct addition reaction between non-decomposed molecules.

Not only equilibrium configurations $A+B$ and $AB$ but a continuous transition from one state to the other is the main goal of computations. Following this way we are able to get the barrier profile of the reaction under consideration. Computationally, it seems quite identical to study the profile by either shifting monomer molecules of the *A+B* complex towards each other, thus contracting the correspondent intermolecular C-C distances, or separating molecules of the *AB* product by elongation of the relevant C-C spacing. A particular dependence of the contact place of the *AB* product on which namely atoms of monomer $C_{60}$ molecules enter the reaction has forced us to prefer the second way. We shall start from the *AB* state and perform a necessary transition to the *A+B* state as a stepwise enlarging of the lengths of intermolecular C-C bonds responsible for the *AB* product formation.

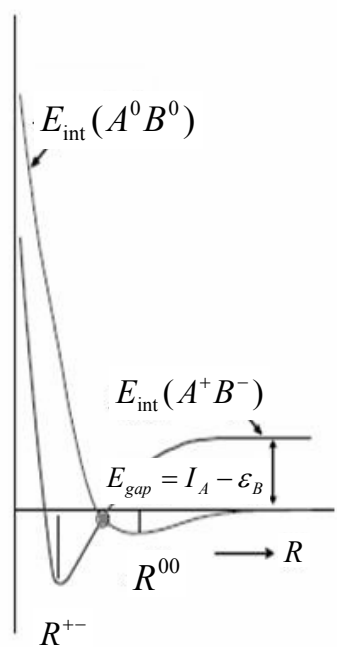

**Figure 1**. Scheme of terms corresponding to the IMI potential of type 1. $(A^0B^0)$ and $(A^+B^-)$ match branches of the terms related to the IMI between neutral molecules and their ions, respectively.

The computational covalent chemistry of the $C_{60}$ molecule is well quantitatively guided by the atomic chemical susceptibility (ACS) map quantitatively representing chemical activity of the molecule atoms [11-13]. The ACS distribution over atoms is presented in figure 2a by different colors that distinguish atoms with different ACS. Among the latter, the most active atoms are shown by light gray. Those are the first targets involved into initial stages of any addition reaction. Consequently, the initial composition of a pair of the $C_{60}$ molecules shown in figure 2a is quite evident. The starting configuration corresponds to $R_{CC}^{st}$ equal to 1.7 Å that corresponds to distances between 1-1` and 2-2` target atoms of group 1. A bound dumbbell-like dimer is formed (figure 2b) after the structure optimization is completed. Two monomers within the dimer are contacted via a typical [2+2] cycloaddition of '66' bonds that form a cyclobutane ring. The latest structure study of the crystalline photopolymers [14] has convincingly exhibited the most abundant component related to [2+2] cycloadduct dimers indeed. The main electronic characteristics of the $(C_{60})_2$ adduct are presented in table 1. A detail comparison with other computational data is given elsewhere [10]. A large negative $E_{cpl}^{tot}$ value undoubtedly evidences that $(C_{60})_2$ dimer is actually a typical *AB* adduct attributed to the $R^{+-}$ minimum on the IMI ground state term in figure 1.

When $R_{CC}^{st} = 3.07$ Å, optimization of the initial structure leads to a weakly bound pair of molecules spaced by $R_{CC}^{fin} = 4.48$ Å. Monomer molecules preserve their initial configurations and, as seen from table 1, form a classical charge transfer complex. The fragment composition of the HOMO and LUMO is cross-partitioned, which means that when the former should be attributed to Mol 2 while the latter – to Mol 1, showing that a charge transfer from Mol 2 to Mol 1 occurs when the complex is photoexcited.

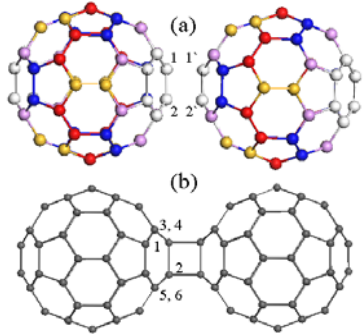

**Figure 2**. *a*. Start composition of the $C_{60}$ + $C_{60}$ composite. *b*. Equilibrium structure of the $(C_{60})_2$ dimer; $R_{CC}^{st} = 1.7$ Å, $R_{CC}^{fin} = 1.55$ Å. All distances correspond to the spacing between 1-1` and 2-2` atoms.

To obtain the barrier profile, two intermolecular C-C distances, namely, 1-1' and 2-2' separations of the [2+2] cycloaddition in figure 2a, were stepwise elongated with a constant increment of 0.05Å during the first stage of elongation from 1.57 to 2.22Å and then of 0.1Å during the second stage. $C_{60}$+CNT and $C_{60}$+graphene composites were similarly treated in the current study.

**Table 1**. Electronic characteristics of the $C_{60}$+$C_{60}$ composite

| Computed quantities (AM1) singlet, UBS HF AM1 singlet state | Monomer $C_{60}$ | $R_{CC}^{st}$ 1.71 E | 3.07 E |
|---|---|---|---|
| Heat of formation[1], $\Delta H$, kcal/mol | 955.56 | 1868.49 | 1910.60 |
| Coupling energy[2], $E_{cpl}^{tot}$, kcal/mol | - | -42.63 | -0.52 |
| Ionization potential[3], $I$, eV | 9.86 (8.74[a]) | 9.87 | 9.87 |
| Electron affinity[3], $\varepsilon$, eV | 2.66 (2.69[b]) | 2.62 | 2.66 |
| Dipole moment, $Db$ | 0.01 | 0.001 | 0.001 |
| Squared spin, (S**2) | 4.92 | 10.96 | 9.87 |
| Total number of effectively non-paired electrons, $N_D$[4] | 9.84 | 21.93 | 19.75 |
| Gained charge to *Mol 1* | - | 0.0 | 0.0 |
| Transferred charge from *Mol 2* | - | 0.0 | 0.0 |
| Symmetry | $C_i$ | $C_{2h}$ | $C_i$ |
| HOMO, fragment compositions, $\eta$ | - | $\eta_{Mol1}$=61.8% $\eta_{Mol2}$=38.1% | $\eta_{Mol1}$=0% $\eta_{Mol2}$=100% |
| LUMO, fragment compositions, $\eta$ | - | $\eta_{Mol1}$=83.9% $\eta_{Mol2}$=15.8% | $\eta_{Mol1}$=100% $\eta_{Mol2}$=0% |

[1] Molecular energies are heats of formation $\Delta H$ determined as $\Delta H = E_{tot} - \sum_A \left( E_{elec}^A + EHEAT^A \right)$. Here $E_{tot} = E_{elec} + E_{nuc}$, while $E_{elec}$ and $E_{nuc}$ are the electron and core energies. $E_{elec}^A$ and $EHEAT^A$ are electron energy and heat of formation of an isolated atom, respectively.
[2] Coupling energy is determined by Eq (1).
[3] Here $I$ and $\varepsilon$ correspond to the energies of HOMO and LUMO, respectively, just inverted by sign. Experimental data for the relevant orbitals are taken from [15] (*a*) and [16] (*b*).
[4] The total number of effectively non-paired electrons, $N_D$, displays the molecular chemical susceptibility of the species [11-13].

Figure 3 exhibits the barrier profile of the $C_{60}$ dimerization in terms of the total coupling energy $E_{cpl}^{tot}$ determined as

$$E_{cpl}^{tot}(R_{CC}) = \Delta H_{dim}(R_{CC}) - 2\Delta H_{mon}^{eq} \qquad (1)$$

where $\Delta H_{dim}(R_{CC})$ and $\Delta H_{mon}^{eq}$ present heats of formation of dimer at the current intermolecular C-C distance and monomer in equilibrium, respectively. This energy is evidently complex by nature since at least two components contribute into the energy value such as the energy of both monomers deformation $E_{def}^{tot}$ and the energy of the covalent coupling $E_{cov}^{tot}$ between the monomers. The former component related to both monomer molecules is determined as

$$E_{def}^{tot}(R_{CC}) = \Delta H_{mon1}(R_{CC}) + \Delta H_{mon2}(R_{CC}) - 2\Delta H_{mon}^{eq} \qquad (2)$$

Here $\Delta H_{mon1}(R_{CC})$ and $\Delta H_{mon2}(R_{CC})$ present heats of formations of one-point-geometry monomer molecules in the structure that correspond to the dimer one at $R_{CC}$ intermolecular distance. The second component $E_{cov}^{tot}$ we determine as

$$E_{cov}^{tot}(R_{CC}) = E_{cpl}(R_{CC}) - E_{def}^{tot}(R_{CC}) \qquad (3)$$

As seen in figure 3, the deformation energy is the largest in the equilibrium dimer and then it steadily decreases when $R_{CC}$ grows remaining positive until becoming zero when monomers are spaced more than 3Å. Similarly, the energy of the covalent coupling, the largest in the equilibrium dimer, then steadily decreases by absolute value being negative and showing a clearly vivid maximum that coincides with that of $E_{cpl}^{tot}(R_{CC})$ after which it is falling by absolute value approaching to zero for largely spaced monomers. Referring to schemes of electronic terms in figure 1, one should accept that this is the energy $E_{cov}^{tot}$ that should be attributed to the *netto* barrier profile. However, energy $E_{cpl}^{tot}(R_{CC})$ as a *brutto* barrier profile will obviously govern the dimerization of fullerene molecules in practice.

### 3. $C_{60}$ oligomers

Since the discovery of magnetism in all-carbon crystals consisting of polymeric layers of covalently bound $C_{60}$ molecules [17], there has been a splash of interest to the polymer structures. Experimentally observed there are three packing modes related to polymer structure of $C_{60}$, namely, one linear mode characteristic for the orthorhombic (*O*) crystalline modification and two planar, or carpet packing modes corresponding to the tetrahedral (*T*) and rhombohedral (*R*) crystal modifications [18].

To construct the related model structures let us first consider the polymerization as a stepwise reaction $(C_{60})_n = (C_{60})_{n-1} + C_{60}$. Each successive step deals with a dyad of the $(C_{60})_{n-1} + C_{60}$ type. As seen from table 1, the dimer ionization potential $I$ and electron affinity $\varepsilon$ practically coincide with those of monomers. The same is characteristic for oligomer $(C_{60})_{n-1}$ for which the related $I$ and $\varepsilon$ are practically the same as for monomer $C_{60}$. Accordingly, the same IMI potential governs the formation of both $C_{60}$ dimer and oligomers and two products, namely $(C_{60})_{n-1} + C_{60}$ charge transfer complex and $(C_{60})_n$ oligomer correspond to equilibrium positions at $R^{00}$ and $R^{+-}$ minima of the IMI term. A coexistence of such two products was experimentally supported in the case of $C_{60}$ trimers in both physical 'dry' [19-21] and chemical 'wet' [22] experiments.

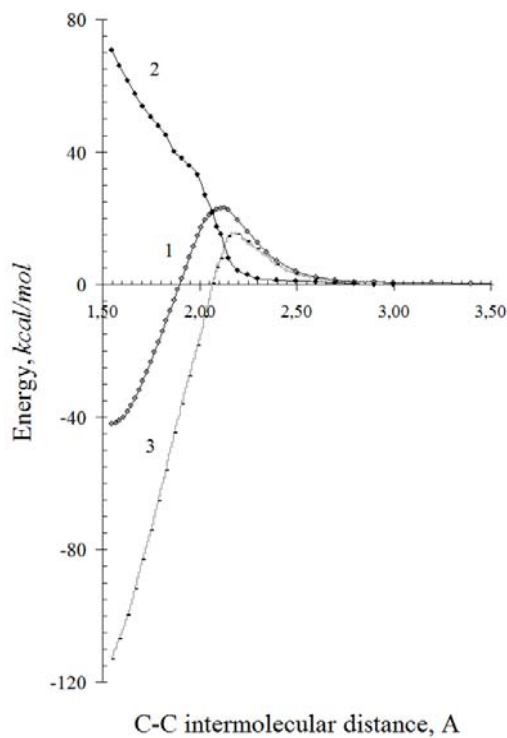

**Figure 3**. Profile of the barrier of $C_{60}$ dimerization. 1. $E_{cpl}^{tot}$; 2. $E_{def}^{tot}$; 3. $E_{cov}^{tot}$.

Following concepts of computational chemistry of fullerenes presented in [11-13], let us consider the trimer formation in the same manner that was applied to fullerene dimer in the previous section. According to the ACS map of dimer $(C_{60})_2$, there are two groups of the high-rank-$N_{DA}$ atoms. The first group combines the most reactive atoms 3, 4, 5, and 6 (see figure 2). Next by reactivity four atoms form two pairs located at the top of both monomers (see figure 4). Enhanced ACS of the first four atoms is evidently connected with a considerable elongation of bonds 2 and 3, which results in weakening the interaction of odd electrons connected with the related atoms. However, in spite of their high chemical reactivity these atoms are non-accessible in due course of the further oligomerization so that top atoms on the right monomer are the most important. Following these ACS indication, a right-angle-triangle trimer ($90^0$-trimer) must be produced. On the other hand, edge atoms located at the end of horizontal axis are characterized by the least values on the ACS map, so that the formation of a linear trimer is highly improbable.

'Wet' chemical experiments [22] dealt with $C_{180}$ species obtained in a 'chemical flask' under convenient chemical conditions and the selected product was afterwards investigated by using STM. Two fractions (A and B in ratio of ~ 5:4) were obtained, the former predominantly (~60%) consisting of $90^0$-trimers while 100% of the latter fraction are presented by cyclic $60^0$- trimers. 'Dry" physical experiments deal with trimers produced under photoilumination of either $C_{60}$ films preliminary deposited on some substrates [19-21] or pristine $C_{60}$ crystal [23]. Only linear three-ball chains were observed in these studies.

Therefore, the wet experiment, as might be expected, obviously supports the preference of $90^0$-trimers predicted by the ACS-guided covalent chemistry of $C_{60}$. Three other trimers, namely, $108^0$-, $120^0$-, and $144^0$-ones observed experimentally within A fraction [22], are of higher energy, which follows from both our and previous [22, 24] calculations. Kinetic conditions may allow for producing these adducts as well whilst with less probability due to energetic unfavorableness. But how can be commented the availability of linear and $60^0$- trimers? As for the former, since from the 'vacuum' quantum chemistry viewpoint their formation is hardly possible, one must suppose the existence of other reasons that may facilitate the trimer production. As turns out, the observation of similar events is not rare and those are usually related to *topochemical reactions* whose occurring is controlled by the reactants packing. Thus, the polymerization of $C_{70}$ fullerene in solid state is explained in terms [25] when "the alignment of molecules and their presumed orientational mobility facilitate polymerization via spatial adjustment of reactive double bonds of neighboring cages". All observed 'dry' polymerization events should be obviously attributed to results of topochemical reactions. Consequently, chemical IMI is responsible only for a part of energetics in this case while molecular packing plays a very important role. The latter will be followed by considerable strain of originated dimers, trimers, and other oligomers, which might, in its turn, considerably redistribute the ACS over molecule atoms.

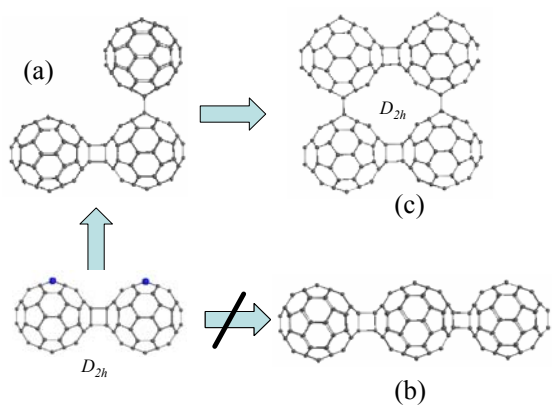

**Figure 4**. Scheme of the stepwise $C_i$-based $C_{60}$ polymerization; favorable (*a, c*) and unfavorable (*b*) prolongations. Equilibrium structures of dimers, trimer, and tetramer. Blue balls on dimer mark high-rank reactive atoms.

The presence of the $60^0$- trimers of fraction B in wet experiments has no relation to preliminary molecules packing and should be explained on the basis of the covalent chemistry of $C_{60}$. However, it cannot be done within the framework of the suggestion that $C_{60}$ is presented in the action by only one $C_{60}$ isomer of $C_i$ symmetry [11]. The ACS map of this isomer well explains a preferable trimerization towards $90^0$-trimers and less profitable one in regards to $108^0$-, $120^0$-, and $144^0$-ones in the framework of the scheme presented in figure 4. Moreover, $D_{2h}$ symmetry of both dimer itself and its monomer is in good correlation with the same symmetry of tetramer and its monomer, so that trimer in figure 4 may be considered as a precursor of tetragonal packed oligomers of higher order as well. As a whole, scheme of polymerization suggested in figure 4 can be considered as a scheme of the $C_i$-based polymerization.

The fact that cyclic $60^0$-trimers form a separate fraction indicates that their difference from trimers of fraction A concerns the origin and kinetics and must be deeply implanted into the monomer structure and properties. Actually, the trimers formation can be understood if $C_i$-symmetric isomer $C_{60}$ is substituted by $D_{3d}$-symmetry one (see a detail discussion on $C_{60}$ isomers in [11]). As shown, there is such an isomer within the $C_{60}$ isomer family that is isoenergetic with $C_i$ one and that may 'go out of the shadow' to provide particular chemical reactions. Let us see how might look a $D_{3d}$-based $C_{60}$ polymerization scheme. As seen in figure 5, changing in monomer symmetry results in another view on the high-rank $N_{DA}$ ACS distribution as well as in another manner of monomer coupling under oligomerization. UBS HF calculations show that the $D_{3d}$-based dimer is of the same energy that $C_i$-based one. The same should be told concerning coupling energies per one [2+2] cycloaddition of $C_i$- and $D_{3d}$-based trimers so that the cyclic trimer is highly stabile. This is consistent with data obtained in the framework of restricted computational schemes [22, 24]. Similarly to the precursor role of the $C_i$-based trimer in the formation of tetragonally packed (TP) higher oligomers, $D_{3d}$-based trimer play the same role for hexagonally packed (HP) ones.

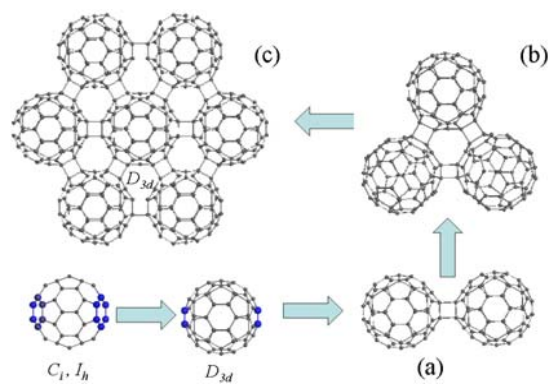

**Figure 5**. Scheme of the stepwise $D_{3d}$-based $C_{60}$ polymerization. Equilibrium structures of dimer, trimer, and heptamer. Blue balls mark high-rank reactive atoms of monomer.

**4. Barrier profile for $C_{60}$-CNT nanobuds**
There have been known a few attempts to synthesize $C_{60}$+CNT complexes that present a single structure in which the fullerenes are covalently bonded to the outer surface of the tubes. For the first time, the covalently bound $C_{60}$ with single-wall CNT (SWCNT) was obtained by means of solid phase mechanochemical reactions [32]. The next time the $C_{60}$+SWCNT complex was synthesized via a microwave induced functionalization approach [33]. It has been used as a component of the photoactive layer in a bulk heterojunction photovoltaic cell. The results indicate that $C_{60}$-decorated SWCNTs are promising additives for performance enhancement of polymer photovoltaic cells. Simultaneously a large investigation has been performed to produce fullerene-functionalized SWNTs, which were termed nanobuds (carbon nanobuds (CNBs) below) and which were selectively synthesized in two different one-step continuous methods, during which fullerenes were formed on iron-catalyst particles together with SWNTs during CO disproportionation [34, 35]. It was suggested that the field-emission characteristics of CNBs might possess advantageous properties compared with SWCNT or fullerenes alone, or in their nonbonded configurations.
     Computational consideration of CNBs has been restricted so far to two publications [34, 36]. The computations were performed in the framework of the density functional theory (DFT) by using periodic boundary conditions (PBCs) in the restricted closed-shell approximation. A few compositions of intermolecular C-C bonds that form the contact zone on sidewall of the tubes have been considered among which [2+2] cycloaddition turned out to be the most efficient. Moving out of PBCs, considering the formation of CNB occurred in due course of a DA reaction in terms of general energetic scheme presented in figure 1, and applying broken symmetry approach, which is based on unrestricted open shell approximation that is more suitable for partially radicalized both fullerene $C_{60}$ and CNTs [11, 37], we present a comprehensive chemical view on the CNBs behavior.

*4.1. Computational synthesis of carbon nanobuds*
Partial radicalization of CNTs is connected with effective unpairing of their odd electrons due to a relative elongation of C-C bonds in comparison with a critical value of 1.395Å under which odd electrons are fully covalently bound forming classical π electron pairs [37]. The effectively unpaired electrons form a pool of molecular chemical susceptibility determined by the total number of the unpaired electrons $N_D$. Distributed over the tube atoms by partial number of effectively unpaired electrons $N_{DA}$, the electron highlights the map of chemical activity of the tube in terms of atomic chemical susceptibility (ACS) $N_{DA}$. Figure 6 presents the ACS distribution over atoms of two (4,4) SWCNTs shown in figure 7. The tubes differ by edge atoms that are non-saturated or empty in figure 7a (tube 1) and terminated by hydrogen atoms in figure 7b (tube 2). The atom numbering of plottings in figure 6 corresponds to that one in the tubes following from the tube caps to their ends.

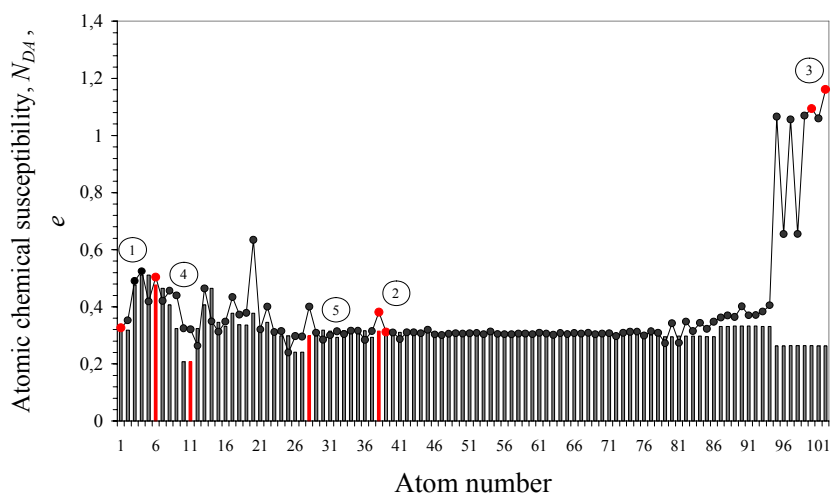

**Figure 6.** Distribution of the **a**tomic chemical susceptibility $N_{DA}$ over atoms of fragments of (4,4) single walled carbon nanotubes with empty (curve with dots) and hydrogen-terminated (histogram) ends [37]. Ringed numbers denote atom pairs subjected to further $C_{60}$ addition (see text).

As seen in figure 6, There are three zones in the ACS distribution related to the tube caps, sidewalls, and ends. According to this, we have chosen three pairs of target atoms on tube 1 (1, 2, and 3), which are shown by red dots on the curve, and two pairs on tube 2 (see pairs 4 and 5 of red bars in histogram). As was discussed in section 2, according to the chemical portrait of the fullerene $C_{60}$, its most active atoms form particularly oriented two hexagons and each pair of them joined by a short C-C bond is the target that first meets any addend. The two features related to CNTs and fullerene molecule make it possible to construct starting configurations of possible CNBs when looking their equilibrium configurations presenting $AB$ products in the scheme in figure 1. Such selected sets of configurations alongside with equilibrium structures of thus obtained CNBs are presented in figure 8.

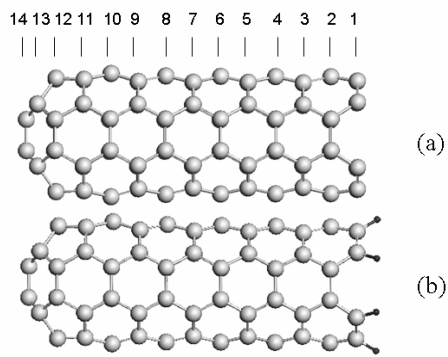

**Figure 7**. Equilibrium structures of (4,4) single-walled carbon nanotube fragments with empty (a) and hydrogen-terminated (b) ends. Figures mark atom's rows.

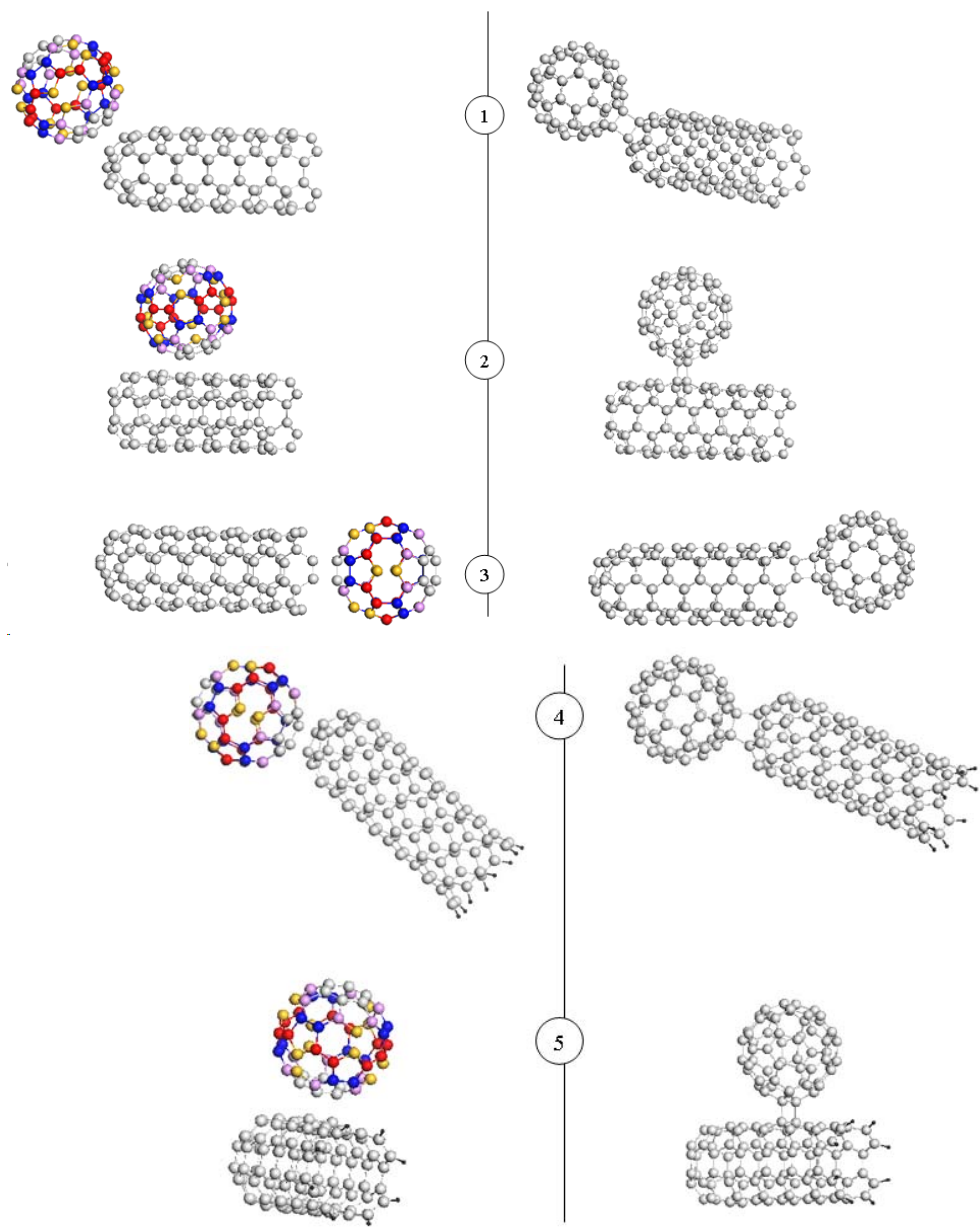

**Figure 8.** Start (left) and equilibrium (right) structures of carbon nanobuds formed by attaching $C_{60}$ to (4,4) single walled carbon nanotube with empty (1, 2, 3) and hydrogen-terminated (4, 5) ends.

Figure 9 presents changing in the ACS maps of both (4,4) SWCNT and fullerene $C_{60}$ after formation of CNB 5. As seen in the figure, the attachment of the fullerene molecule to the sidewall of the tube causes only local changing that concerns atoms participating in the formation of the [2+2] cycloaddition. The other part of the atomic activity distribution of the tube retains non-perturbed. This finding evidently favors a multiple attachment of fullerenes to the tube in a superposition manner. Oppositely, the fullerene ACS map changes considerably indicating a significant redistribution of the atomic chemical activity over the molecule atoms after attachment. Red dots on plotted curve in figure 9b highlight new most active atoms prepared for the next reaction events.

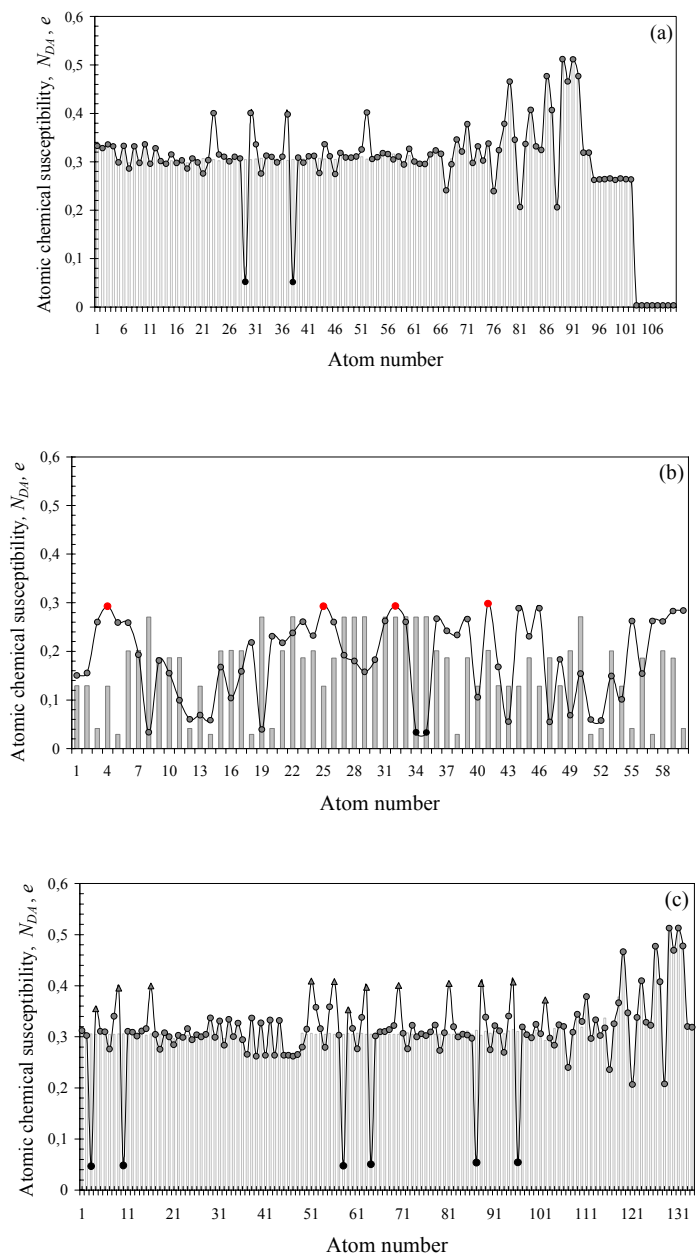

**Figure 9**. Evolution of atomic chemical susceptibility under CNB formation. *a*. tube 2; *b*. fullerene $C_{60}$; *c*. tube 3 (see text). Histograms present data for the pristine species. Curve with dots plot data related to the formed CNBs.

To check a high tolerance of tube body to a multiple attachment of the fullerene molecules, two and three $C_{60}$ were attached to an elongated (4,4) SWCNT (tube 3) forming CNB 6 and CNB 7 (see figure 10). As shown in [37], the tube elongation causes the elongation of the sidewall zone in the ACS map only and does not touch either cap or end atoms regions. That is why conditions for the formation of CNDs 5, 6, and 7 are identical. All $C_{60}$ molecules are covalently coupled with the tube body via [2+2] cycloadditions. Changing in the ACS distribution related to the tube is shown in figure 9c. A clearly seen superposition of the three attachments is perfectly exhibited by the map indicating that practically countless number of fullerene molecules can be attached to SWCNTs long enough.

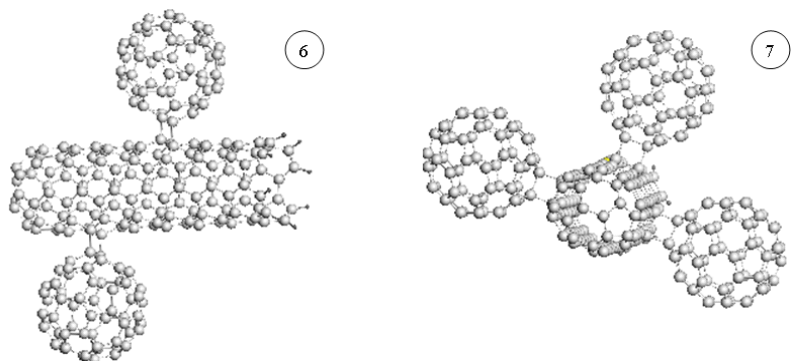

**Figure 10**. Equilibrium structures of CNBs related to double (CNB 6) and triple (CNB 7) attachments of $C_{60}$ to the sidewall of (4,4) single walled carbon nanotube with hydrogen-terminated ends.

Figure 11 presents a collection of the studied CNBs with indication of most active fullerene's atoms that may continue the nanobuds chemical modification. A detail analysis of the presented structures shows that the intermolecular junctions as [2+2] cycloaddition are formed only in the case when fullerene is covalently coupled with the tubes bodies. The junctions in CNBs 1 and 3 are not typical for the cycloaddition in spite of four-atom membership. Color highlighting of the most active atoms of attached fullerenes presented for various CNB configurations in figure 11 straightly points to a similar behavior of the attached molecule. Each of them has two atom pairs numbered 1 in the equatorial plane added by a pair 2 of atoms in the vicinity of [2+2] contacts similarly to the situation described for $C_{60}$ oligomer in section 3. Analogously the case, atoms of pair 2 are accessible only for very small addends thus putting atoms of pairs 1 on the first place in the continuation of the CNBs chemical modification via subsequent attachments to the fullerene, the best suitable for, say, application related to photovoltaic cells.

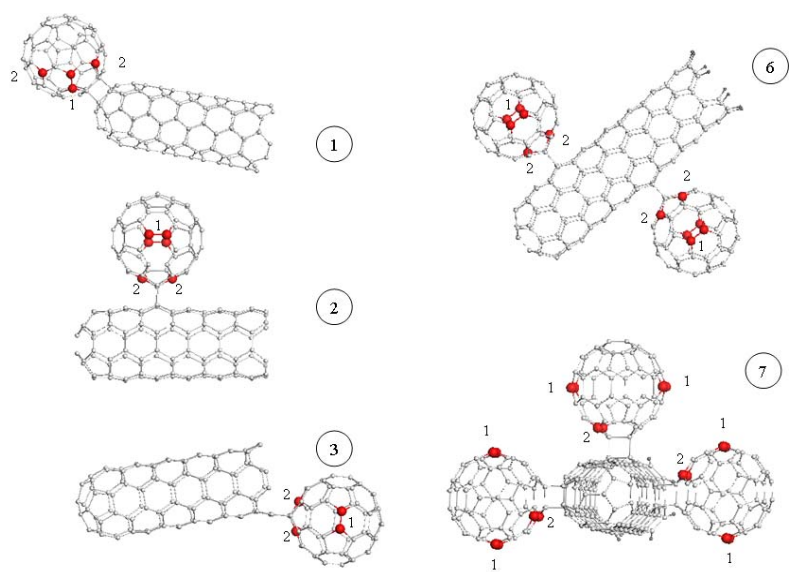

**Figure 11**. Equilibrium structures of CNBs. Red balls indicate $C_{60}$ atoms with high-rank $N_{DA}$ values.

*4.2. Energetic parameters and reaction barrier*

Energetic characteristics related to the obtained equilibrium CNBs are presented in table 2. The coupling energy $E_{cpl}^{tot}$ is determined as

$$E_{cpl}^{tot} = \Delta H_{CNB} - \Delta H_{CNT} - \Delta H_{C_{60}}. \tag{4}$$

Here $\Delta H_{CNB}$, $\Delta H_{CNT}$, and $\Delta H_{C_{60}}$ present heats of formation of the equilibrium structures of CNB, (4,4) SWCNT, and $C_{60}$, respectively.

**Table 2.** Energetic characteristics of [$C_{60}$-(4,4)] carbon nanobuds, *kcal/mol*

| Nanobuds[1] | $E_{cpl}^{tot}$ | $E_{def}^{tot}$ | $E_{defCNT}$ | $E_{defC_{60}}$ | $E_{cov}^{tot}$ |
|---|---|---|---|---|---|
| 1 (cap) | -36,33 | 51,16 | 10,62 | 40,53 | -87,48 |
| 2 (wall) | -3,38 | 59,64 | 24,64 | 35 | -63,02 |
| 3 (end) | -86,65 | 47,65 | 8,25 | 39,4 | -134,31 |
| 4 (cap) | 3,09 | 114,38 | 62,76 | 51,62 | -111,29 |
| 5 (wall) | -4,26 | 74,33 | 39,26 | 35,07 | -78,59 |
| 6 (wall) | -8,21 (-4,10)[2] | 155,62 | 85,33 (42,66)[2] | 70,29 (35,15)[2] | -163,83 (-81,92)[2] |
| 7 (wall) | -11,02 (-3,67)[2] | 221,64 | 116,59 (38,86)[2] | 105,05 (35,02)[2] | -232,66 (-79,55)[2] |

[1] Figures number CNBs as in figures 8 and 10
[2] Data per one attached $C_{60}$ molecule.

As discussed in section 2, we suggest that the total coupling energy reflects two processes that accompany the fullerene attachment to the tube, namely, the deformation of both CNB components and their covalent coupling. The energy caused by deformation can be determined as

$$E_{def}^{tot} = E_{defCNT} + E_{defC_{60}}, \tag{5}$$

where

$$E_{defCNT} = \Delta H_{CNT}^{CNB} - \Delta H_{CNT} \quad \text{and} \quad E_{defC_{60}} = \Delta H_{C_{60}}^{CNB} - \Delta H_{C_{60}}. \tag{6}$$

Here $\Delta H_{CNT}^{CNB}$ and $\Delta H_{C_{60}}^{CNB}$ present heats of formation of one-point-geometry configurations of the SWCNT and fullerene components of the equilibrium configurations of the studied CNBs. Accordingly, the chemical contribution into the coupling energy is determined as

$$E_{cov}^{tot} = E_{cpl}^{tot} - E_{def}^{tot}. \tag{7}$$

Analyzing data in table 2, two distinct features should be emphasized. The first concerns the difference in the behavior of CNBs, fullerene components of which are attached to either cap or end atoms of the tube. The second is related to a close similarity in the properties of CNBs with fullerene attached to the tube sidewall. One can see practically identical characteristics for these CNBs based on hydrogen-terminated tubes alongside with a rather small deviation of the data from that related to the tube with empty ends. Additionally, as seen in the table, the empty end of the tube, once the most active according to the ACS map in figure 6, provide the fullerene attachment with the biggest coupling energy and the smallest deformation energy related to the tube. It means that at any contact of such tube with fullerene, the first attachment occurs on the tube end. The next events will take place at the tube cap. The coupling energy decreases by 2.4 times but the deformation energy slightly increases. After these two events, comes a turn of the tube sidewall, but the coupling energy decreases by ~26 times when the deformation energy increases three times. The three events are quite superpositional as can be seen in figure 12. Each addition concerns a strictly local area so that highly active attachments in the end and cap regions should not prevent from covering the main tube body by multiply attached fullerenes.

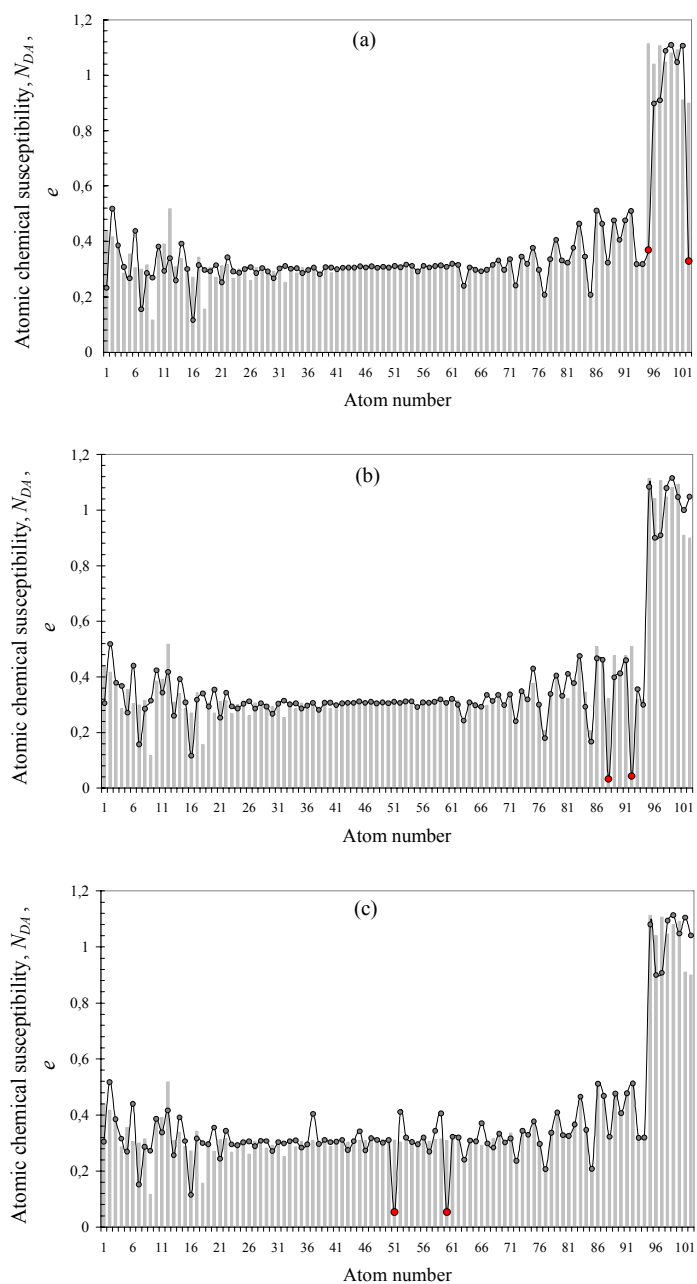

**Figure 12**. The evolution of atomic chemical susceptibility of tube 1 under CNB formation when $C_{60}$ is attached to the tube empty end (a), cap (b), and sidewall (c). Histogram presents data for the pristine tube. Curve with dots plot data related to the formed CNBs 1, 2, and 3, respectively. Red dots mark atoms, to which fullerene is attached.

Obviously, end- and cap attachments are quite exclusive for the CNBs, the main configurations of which are formed by fullerene molecules attached to the tube extended sidewall. The configuration of CNB 5 with a single attachment of fullerene has been chosen for determining the reaction barrier for this most typical case. Figure 13 presents the dependence of the basic energetic characteristics of the CNBs on the C-C intermolecular distance. As in the case of fullerene dimers discussed in section 2, the barrier energy computation has been started from the equilibrium configuration of CNB 5 followed by a stepwise elongation of two outer C-C bonds that provide intermolecular contact via [2+2] cycloaddition. A deep parallelism in the behavior of both singly bound fullerene molecules to the tube body and a single molecule in the case of multiple attachments provides a good reason to expect the same parallelism between the energy dependencies of these molecules as well.

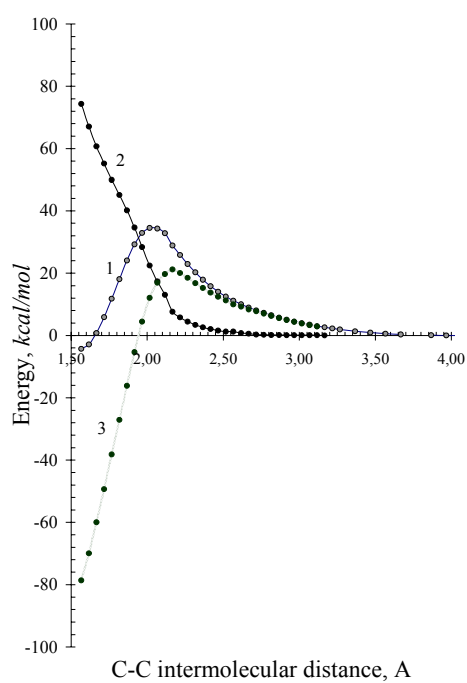

**Figure 13.** Profile of the barrier of the [$C_{60}$-(4,4)] CNB formation. 1. $E_{cpl}^{tot}$; 2. $E_{def}^{tot}$; 3. $E_{cov}^{tot}$.

Three plottings in figure 13 behave quite similarly as in the case of $C_{60}$ dimer (see figure 3), which could be expected since the contact zones in both cases are presented by similar [2+2] cycloadditions. The coupling energy $E_{cpl}^{tot}$ can evidently be divided into $E_{def}^{tot}$ and $E_{cov}^{tot}$ components of the same type as those related to $C_{60}$ dimer. However, the difference in numerical values of the two components at starting point results in much shallower minimum of $E_{cpl}^{tot}$ in the case of CNB and, thus, less barrier for the CNB decomposition.

## 5. Barrier profile for $C_{60}$-graphene composites
In contrast to fullerene oligomers and carbon nanobuds, the latter known now not only for $C_{60}$ [32-35] but for $C_{70}$ as well [35], no indication of the existence of chemically bound fullerene-graphene compositions has been so far obtained. It is difficult to believe that no attempts to produce these intriguing compositions have been undertaken. Serious reasons appear to make not possible to achieve the goal. Computations performed by Wu and Zeng [38] were first to lift the veil above the feature. They have shown that the reaction of covalent addition of $C_{60}$ to graphene basal plane is endothermic and requires a considerable amount of energy oppositely to $C_{60}$ dimer and CNBs discussed in sections 2 and 4. However, the mentioned computations have been carried out in a standard configuration of the spin-restricted PBC DFT approach in spite of the fact pointed by the authors themselves that local and semilocal functionals in DFT generally give poor description of weak interaction. Similarly insufficient is response of the technique to the interaction weakening, which takes place between odd electrons of graphene [37]. Thus, test calculations of the authors within a spin-unrestricted method could not show any difference from the spin-restricted ones due to overpressing the configurational part of the functionals. Similar test within the framework of the Hartree-Fock approach results in 23% (or 641.6 kcal/mol by absolute value) lowering of the total energy of the (9,9) nanographene (($n_a$, $n_z$)nomenclature ($n_a$, and $n_z$ number benzoid units along armchair and zigzag directions) follows the suggested in [39]), which was chosen as supercell in the PCB DFT computations [38], when going from RHF to UHF approach.

Describing possibility to arrange periodic graphene nanobuds (GNBs) [38], the author have concentrated their attention on the basal plane of (9,9) nanographene leaving aside the sheet edges as well as supposing homogeneous chemical activity of carbon atoms through over the sheet. It is actually not the case since the chemical activity distribution over graphene atoms is highly inhomogeneous and its edges can be extremely reactive [1, 40, 41] thus playing the main role in providing covalent coupling of any addend including carbon nanotubes [1] and $C_{60}$ [42]. This is particularly important when elaborating technology of producing graphene-based nanocarbon composites in solution. To eliminate features related to definite peculiarities in the interaction of graphene with fullerene $C_{60}$, we have performed a computational synthesis of [$C_{60}$-(5,5)] GNBs varying the place of contact of the molecule with (5,5) nanographene. Presented below concerns the formation of *AB* product in terms of the scheme shown in figure 1. The product is formed when the starting intermolecular C-C distances are less than 2.2Å. At longer distances, the pair of $C_{60}$ and nanographene forms a classical charge transfer complex where graphene's atoms contribute into the HOMO while $C_{60}$'s atoms govern the LUMO, which causes the charge transfer from graphene to fullerene under photoexcitation.

*5.1. Computational synthesis of graphene nanobuds*
Since the length of C-C bonds in graphene noticeably exceeds the critical value at which a complete covalent bonding of the relevant odd electrons is terminated [11], odd electrons of graphene become effectively unpaired thus providing quite valuable molecular chemical susceptibility $N_D$ of the graphene molecule and atomic chemical susceptibility (ACS) $N_{DA}$ related to each atom. Distributed over the graphene atoms, $N_{DA}$ maps the chemical activity of the molecule atoms. Figure 14 presents the ACS distributions over atoms of (5,5) nanographene under conditions when the sheet edges are either non-terminated (empty) (a) or hydrogen-terminated by one (single-H) (b) and two (double-H) (c) atoms per one carbon. Color pictures present 'chemical portraits' of the three molecules in real space while plotting in figure 14d discloses the ACS distributions by absolute values. The total number of unpaired electrons $N_D$ constitutes 31.04 e, 14.57 e, and 12.23 e in the three cases, respectively. As seen in the figure, the portraits diverge considerably exhibiting the difference in both molecular and atomic chemical activity making the three molecules absolutely different in regards to the same chemical reactions. Non-terminated sheet is the most reactive. Then follow single-H- and double-H terminated ones, the latter is the least active with respect to the total molecular susceptibility. But as seen in figure 14c (right) and than confirmed in figure 14d, the reactivity of atoms with maximum ACS values in the latter case is higher than for single-H terminated sample. The feature highlights very important question to be answered before any additive reaction aimed at producing a wished composite starts: what is the atomic composition in the edge zone of nanographene samples under study? In the case of $C_{60}$, its addition to the graphene sheet will obviously occur quite differently depending on particular-edge sample and the place of contact to the latter. Let us consider possible situations expected for the above cases basing on the relevant ACS maps.

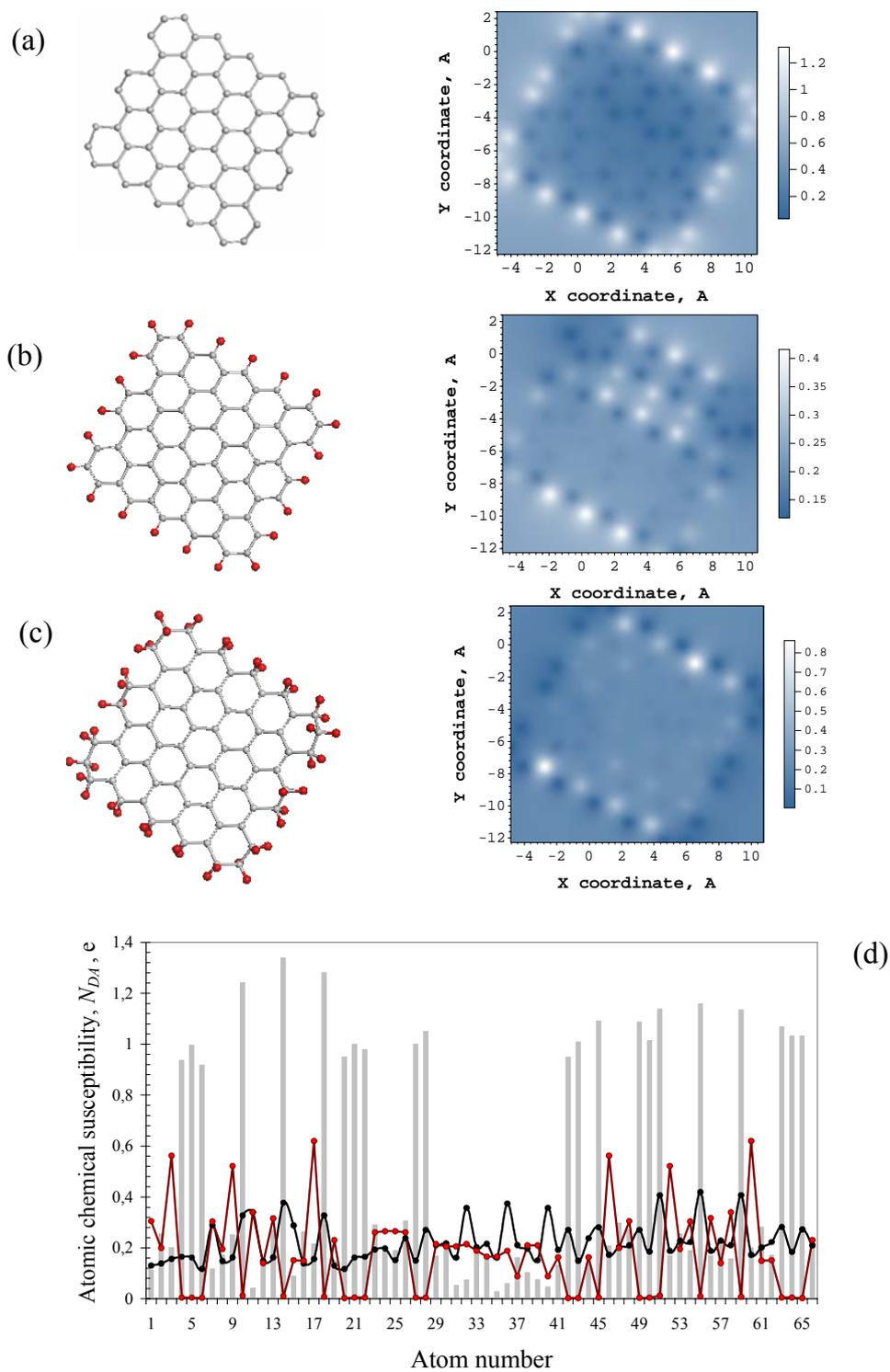

**Figure 14**. Equilibrium structures (left) and chemical portraits (right) of (5,5) nanographene with empty (a), and hydrogen-terminated edge atoms by one (b) and two (c) terminators per carbon atom. Distribution of the chemical susceptibility over nanographene atoms (d) in case a (light gray histogram), b (black curve with dots) and c (dark red curve with dots).

*5.1.1. Deposition of $C_{60}$ on nanographene with non-terminated edges.* According to the ACS map in figure 14a, atoms of zigzag empty edges are the first targets for the fullerene addition. Approaching to the edge, fullerene molecule will orient itself in such a way that provides the closure between its most reactive atoms of group 1 with two zigzag carbon atoms as shown for the relevant GNB 1 in figure 15. Equilibrium structure of the formed GNB is shown next to the start configuration alongside with the ACS maps related to the GNB. Energetic parameters are presented in table 3. We shall refer so far to the total coupling energy $E_{cpl}^{tot}$ only leaving the discussion of other quantities to the next section. The coupling energy is determined as

$$E_{cpl}^{tot} = \Delta H_{GNB} - \Delta H_{gr} - \Delta H_{C_{60}}, \tag{8}$$

where $\Delta H_{GNB}$, $\Delta H_{gr}$, and $\Delta H_{C_{60}}$ determine heats of formation of the considered GNB, graphene itself, and fullerene $C_{60}$, respectively. As seen in the table, the GNB formation is accompanied by high coupling energy whose negative sign points to an energetically favorable process.

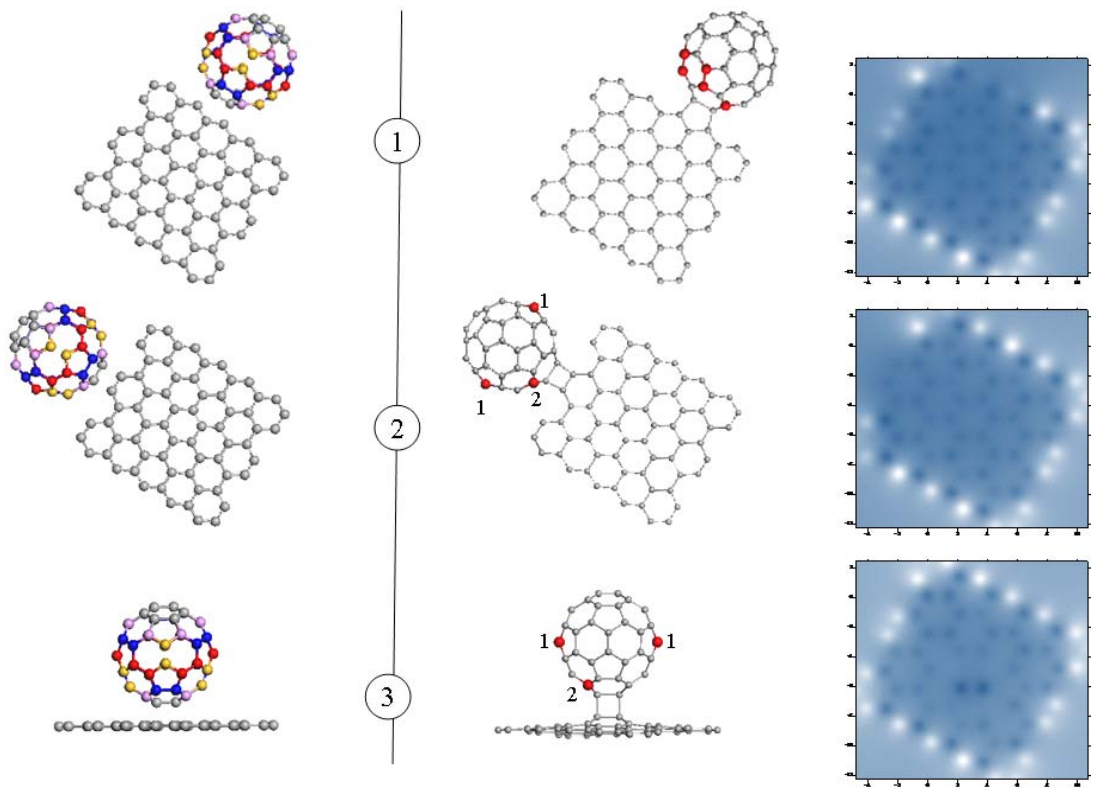

**Figure 15.** [$C_{60}$-(5,5)] graphene nanobuds formed by attaching $C_{60}$ fullerene to zigzag (1) and armchair (2) edge atoms as well as to the basal plane (3) of (5,5) nanographene. Start (left) and equilibrium (right) configurations with real-space ACS maps of the latter. Red balls on equilibrium structures point to fullerene atoms with the highest ACS values. Small number 1 points to pairs of atoms of the same ACS joint by 1,2- connection 'via short bond' [43] while number 2 is related to a pair of atoms joint by 1,4-connection 'via space".

Looking at the ACS map of GNB 1 in figure 15, one can conclude that the GNB formation causes rather local changes in the map. Two brightly shining zigzag atoms in the left upper corner of the map in figure 14a are substituted with two dark spots in figure 15 while retaining ACS of the remainder atoms practically no altered. This conclusion is justified by the ACS plottings presented in figure 16. Therefore, from the ACS viewpoint, the response of both CNTs considered in the previous section and nanographene to the nanobud formation is quite similar. Similar as well is the behavior of the attached fullerene that remains still chemically reactive. Its ACS map considerably changes after addition revealing new target atoms shown by red balls on equilibrium structures.

**Table 3.** Energetic characteristics of [$C_{60}$-(5,5)] graphene nanobuds, *kcal/mol*

| Composite[1] | $E_{cpl}^{tot}$ | $E_{def}^{tot}$ | $E_{defgr}$ | $E_{defC_{60}}$ | $E_{chem}^{tot}$ |
|---|---|---|---|---|---|
| 1zg | -128,45 | 70,56 | 20,7 | 49,86 | -199,01 |
| 2ach | -123 | 52,67 | 13,37 | 39,3 | -175,67 |
| 3b | -12,11 | 86,8 | 53,76 | 33,04 | -98,91 |
| 4zgH1 | -52,74 | 109,89 | 69,08 | 40,81 | -162,63 |
| 5achH1 | -70.17 | 91.53 | 58.10 | 33.44 | -161.71 |
| 6bH1 | -27,38 | 96,63 | 63,48 | 33,25 | -124,11 |
| 7bH2 | 1,82 | 62,24 | 28,97 | 33,27 | -60,42 |
| 8bH1[2] | 32 | 88,98 | 55,75 | 33,23 | -56,98 |

[1] Figures correspond to GNB's numbers in figures 15, 18, and 19.
[2] Data for [$C_{60}$-(9,8)] GNB.

Graphene nanobud 2 formed by attaching fullerene $C_{60}$ to the armchair edge of the graphene sheet is shown in figure 15. The coupling energy $E_{cpl}^{tot}$ that accompanies its formation is big by value and negative (see table 3) just pointing to the energy gain when the product is formed. Analysis of the ACS map in figure 15 and its value in plotting in figure 16b evidences a local perturbation in the odd electronic state of the graphene caused by the fullerene attachment. In contrast to non-regularly localized the most active fullerene atoms designated for the next chemical attack in GNB 1, in the case of GNB 2 there two pairs of ACS-equivalent atoms in the equatorial plane of the fullerene (the situation is typical for the 1,2- addition algorithm of the fullerene derivative formation [11]) marked by 1 and one pairs of atoms connected via 1,4 junction that always is highlighted when a quite bulky addend is attached the fullerene core. As previously, one can judge that the fullerene attachment causes a rather local effect on the ACS distribution.

Placing $C_{60}$ over basal plane of the graphene results in the formation of GNB 3, both starting and equilibrium structures of which is shown in figure 15. Two dark spots on the ACS map of GNB 3 reveal the contact place while the state of the remainder atoms is not vividly distorted. Plotting presented in figure 16c supports this conclusion. Transformation of the ACS map of fullerene is similar to those observed for $C_{60}$ dimer and CNBs discussed previously. The formation of this nanobud is still energetically favorable, though with less coupling energy (table 3).

Summarizing the data obtained for nanographene with empty edges, one should conclude that the sample is highly chemically active, which makes it possible to produce GNBs of different configurations depending on a particular place of contact. Moreover, the configuration of the contact places is also quite different. As seen in figure 15, the junction in GNB 1 involves five carbon atoms (two from the fullerene side and three from the graphene) and is of pentagon shape. The contact zone of GBD 2 is formed by four atoms but it is not a typical [2+2] junction since support atoms 1 and 2 from the graphene side (see figure 2b) have three neighbor atoms instead of four for the support atoms of fullerene. A classical [2+2] junction occurs for GBD 3 only.

Local character of the graphene ACS map perturbation provides the formation of not only mono-$C_{60}$ GNBs considered above, but multiple-$C_{60}$ GNBs looking like one of numerous possible examples presented in figure 17a. In its turn, attached fullerenes may serve as centers for further chemical modification of the GBDs by the formation of branched chains of different configurations (see figure 17b). The variety of possibilities to form different nanobuds based on the same addend is evidently, not a prerogative of fullerene. The same situation will occur with any other addend, once covalently bound with graphene substrate, since it is caused by a sharp variety of the atomic chemical susceptibility over the graphene sheet, on one hand, and the local character of the graphene sheet perturbation caused by each addition, on the other. This feature would obviously influence, say, the local electric response, which will vary depending on places of contact of the same addend, thus putting a serious problem concerning reliability of graphene application as selective sensors of different adsorbates [44].

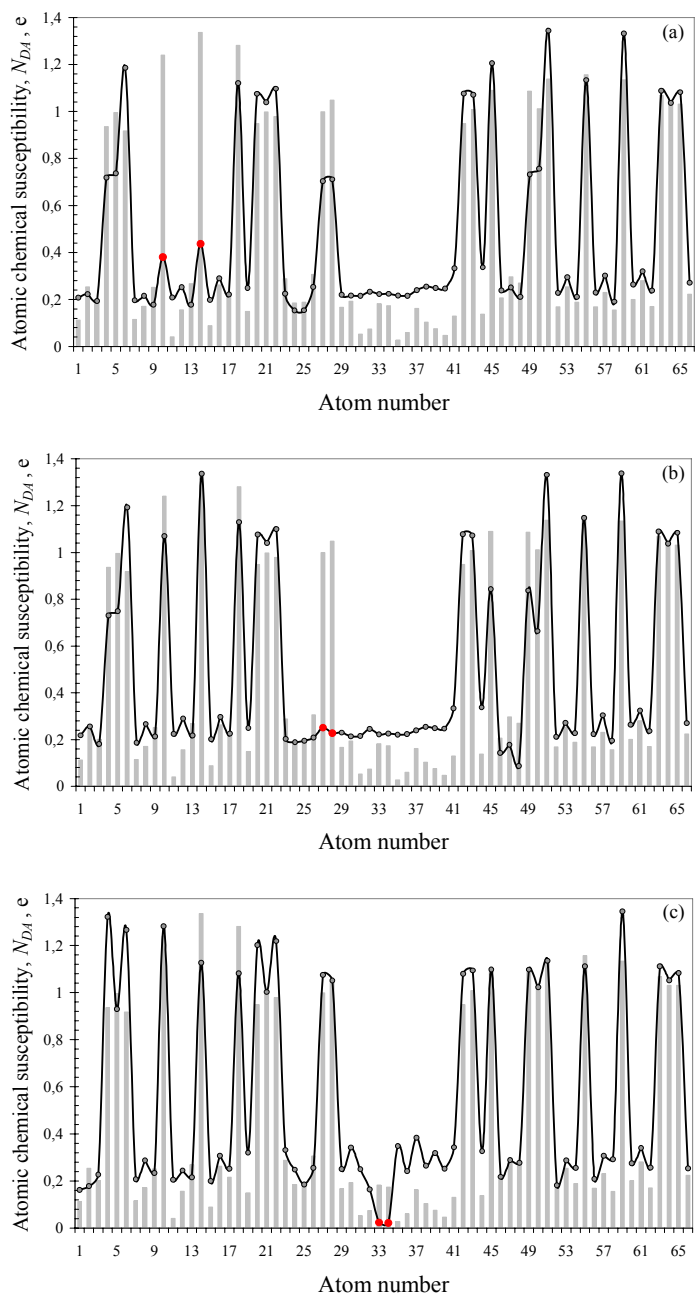

**Figure 16.** Atomic chemical susceptibility distribution over graphene atoms (curves with dots) in GNB 1 (a), GNB 2 (b), and GNB 3 (c). Red spots mark graphene atoms to which $C_{60}$ is covalently attached. Light gray histograms present ACS distribution for the pristine empty-edge (5,5) nanographene.

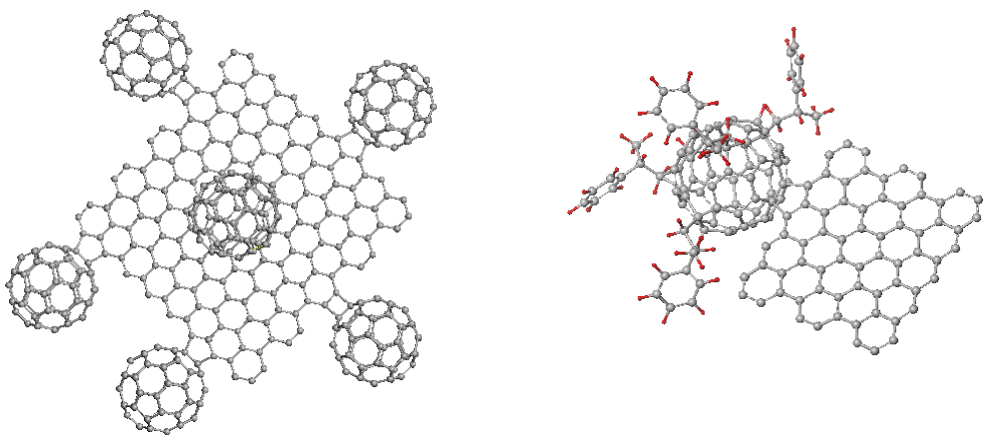

**Figure 17**. *a*. A multiple-$C_{60}$ graphene nanobud. *b*. Graphene nanobud with $C_{60}$-tetrastyrene.

*5.1.2. Deposition of $C_{60}$ on nanographene with hydrogen-terminated edges.* Figure 18 presents GNBs that can be formed in this case. As seen in the figure, the GNBs behavior is absolutely similar to that described in the previous section concerning the local character of the graphene perturbation, the dependence on the place of contact, and the contact zone configurations. The difference is provided by a quantitative difference in the ACS values (see figure 14d). In the case of double-H terminated graphene, the activity of zigzag and armchair edges is fully suppressed so that only contacts on the basal plane participate in the GNBs formation. As previously, the contact zones of GNBs 4 and 5 are not explicitly [2+2] cycloaddition junctions while those of GNBs 6 and 7 belong to the latter.

Changes in the chemical activity of the edge atoms greatly influence energetic parameters of the covalent bonding as seen in table 3. The coupling energies $E_{cpl}^{tot}$ related to the addition to zigzag and armchair edges of single-H terminated graphene decrease more than twice by value. At the same time addition to the basal plane is accompanied by more than twice increase in the value. The double termination of the graphene edge atoms deprives them of noticeable chemical reactivity leaving the only chance for the covalent bonding with carbon atoms located on basal plane. As seen in figure 18, such bonding is formed indeed but needs for its completion ~ 2 kcal/mol.

Graphene nanobud 8 formed by $C_{60}$ covalently coupled with atoms on the basal plane of single-H terminated (9,8) nanographene shown in figure 19 completes the list of studied GNBs. A typical [2+2] cycloaddition forms the contact zone. This composition presents in our opinion the situation that should be often met in real practice. Nanographenes of *nm* in size with definitely terminated edge atoms might be expected as the most abundant part of the graphene solutions [45]. Nanobuds formed at the edges of the sheet behave quite similarly those formed on (5,5) sheet discussed above. However, the formation of GNB 8 highlights a new feature since the coupling energy is high by value but positive, which means that the GNB formation is not energetically profitable in contrast to GNB 6. To elucidate possible reasons for such behavior let us examine energetic characteristics of GNB 8 in details.

## 5.2. Energetic parameters and reaction barrier

As was discussed in sections 2 and 4, it is quite reasonable to present the total coupling energy $E_{cpl}^{tot}$ related to GNBs consisting of two components, namely, $E_{def}^{tot}$ and $E_{chem}^{tot}$ that take the form

$$E_{def}^{tot} = E_{defgr} + E_{defC_{60}}, \tag{9}$$

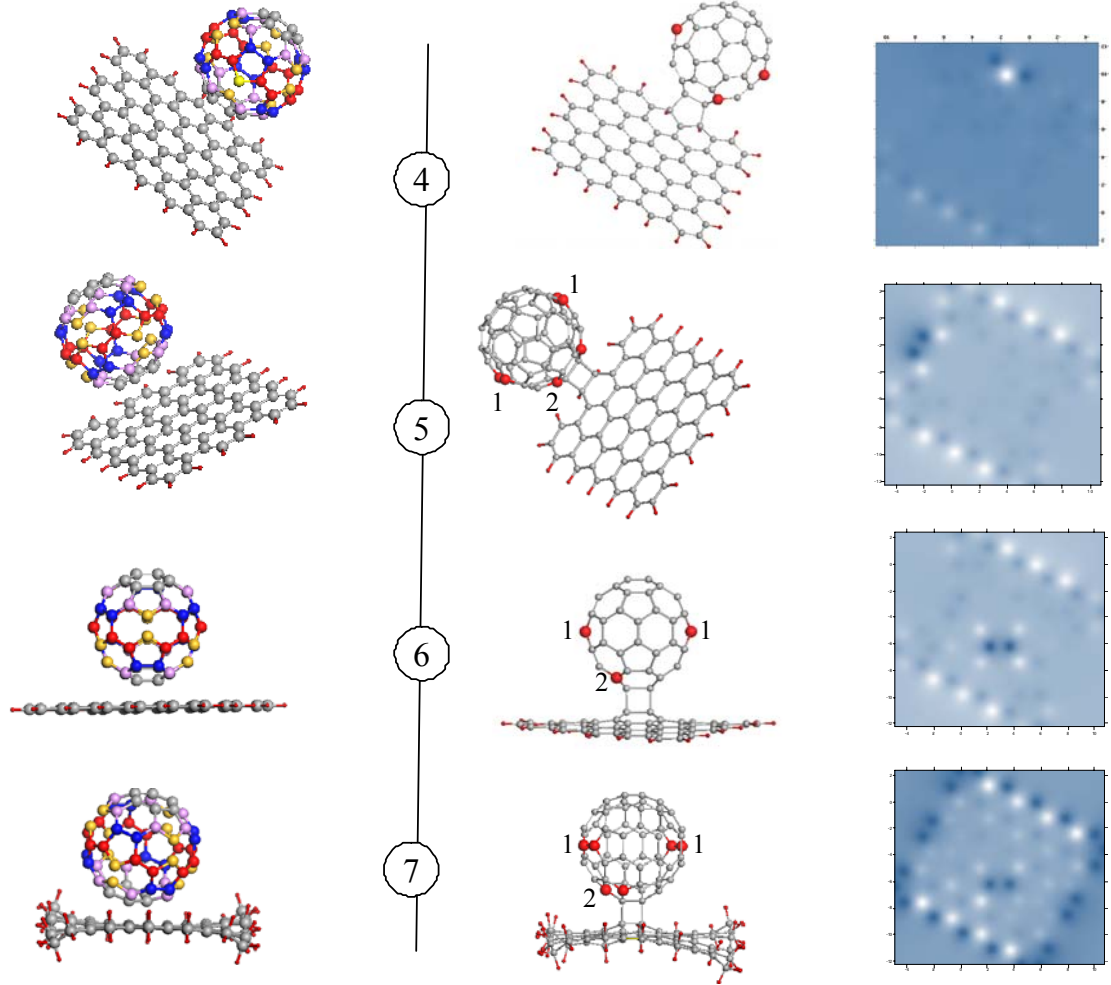

**Figure 18.** [$C_{60}$-(5,5)] graphene nanobuds formed by attaching $C_{60}$ fullerene to zigzag (4) and armchair (5) edge atoms as well as to the basal plane (6) and (7) in the case of single-H (4-6) and double-H (7) terminated (5,5) nanographene. Start (left) and equilibrium (right) configurations with real-space ACS maps of the latter. Red balls on equilibrium structures point to fullerene atoms with the highest ACS values. Small number 1 points to pairs of atoms of the same ACS joint by 1,2- connection 'via short bond' [43] while number 2 is related to a pair of atoms joint by 1,4-connection 'via space".

where

$$E_{defgr} = \Delta H_{gr}^{GNB} - \Delta H_{gr} \quad \text{and} \quad E_{defC_{60}} = \Delta H_{C_{60}}^{GNB} - \Delta H_{C_{60}}. \qquad (10)$$

Here $\Delta H_{gr}^{GNB}$ and $\Delta H_{C_{60}}^{GNB}$ present heats of formation of one-point-geometry configurations of the graphene and fullerene components of the equilibrium configurations of the studied GNB. Accordingly, the chemical contribution into the coupling energy can be determined following Eq.(7). Figure 20 presents the dependence of $E_{cpl}^{tot}$, $E_{def}^{tot}$ and $E_{chem}^{tot}$ on the intermolecular 1-1` and 2-2` C-C distances (see figure 2a) for GNB 8. The three plottings in the figure are generally similar to those presented in figure 3 and figure 13 in regards $C_{60}$ dimer and CNB. This is obviously resulted from the similarity of atomic structure of the contact zones formed in all the considered cases by [2+2] cycloadditions. The coupling energy $E_{cpl}^{tot}$ can evidently be divided into $E_{def}^{tot}$ and $E_{cov}^{tot}$ components of the same type as those related to the previous two cases. However, the difference in numerical values of the two components at start point results in considering lifting of the $E_{cpl}^{tot}$ minimum for GNB 8 moving it into positive energy region. Consequently the barrier energy $E_{barr}^{GNB}$ lowers up to 22.7 kcal/mol. Analyzing data presented in table 3, one can conclude that the observed change in the coupling energy is mainly caused by a drastic lowering of the covalent contribution into $E_{cpl}^{tot}$ in this case.

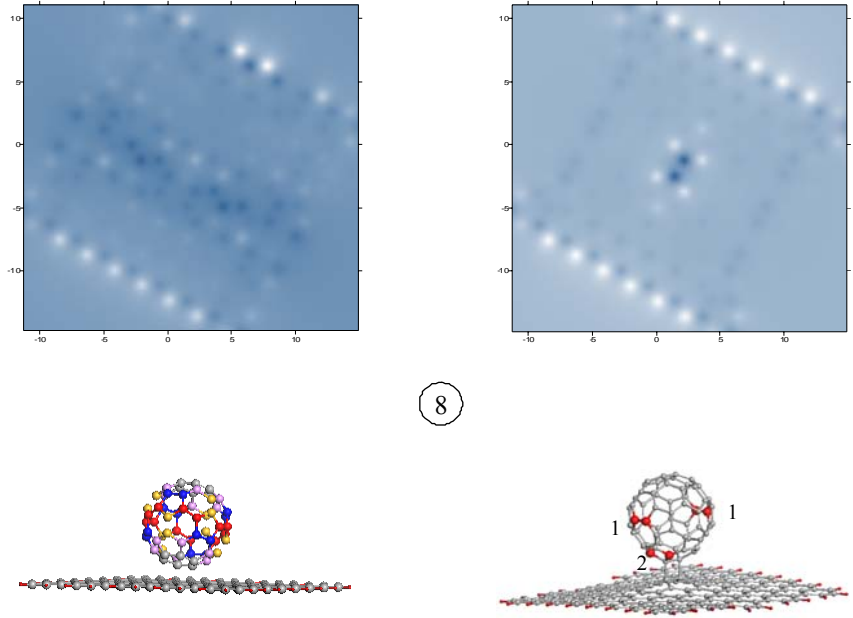

**Figure 19.** [$C_{60}$-(9,8)] graphene nanobud formed by attaching $C_{60}$ fullerene to the basal plane of (9,8) nanographene. Start (left) and equilibrium (right) configurations with ACS maps (above) related to initial and perturbed graphene. Red balls on equilibrium structures point to fullerene atoms with the highest ACS values. Small umber 1 points to pairs of atoms of the same ACS joint by 1,2- connection 'via short bond' [43] while number 2 related to a pair of atoms joint by 1,4-connection 'via space".

## 6. Discussion and conclusive remarks

The studied nanobuds, including $C_{60}$ oligomers, CNBs, and GNBs, are resulted from a covalent pair-pair bonding between the components, where each of the latter delegates a pair of the most chemically active atoms to form intermolecular junctions. The relevant atom pairs have been selected among the others following the indication of the $N_{DA}$ high-rank atoms on the corresponding ACS maps. A high non-homogeneity in the ACS distribution over atoms of SWNT and graphene, which divides their space into three regions, namely: cap, end, and sidewall of CNT as well as zigzag and armchair edges and basal plane of graphene, forms the grounds for the dependence of the formed nanobuds from the place of $C_{60}$ location and from the state-of-art termination of both end and edge atoms. Consequently, the formed nanobuds present a rather complicated set of covalently bound composites that differ by both coupling energies and the structure of intermolecular junctions. Thus, only $C_{60}$ dimer, sidewall CBDs, and basal-plane GNBs can be characterized by the [2+2] cycloaddition as alike intermolecular junction. Table 4 summarizes data covering the three studied cases.

As seen in table 4, structural characteristics concerning the [2+2] cycloadditions are quite alike and practically identical within CNB and GNB groups while the coupling energies differ quite considerably. It is difficult to connect the difference with rather small one concerning $N_{DA}$ values. Therefore, in contrast to a close similarity in both coupling and barrier energies for the studied nanobuds that might be expected in the cases, we see a large variety of properties. It is hidden from view now what is the reason for the observed peculiarity. Obviously, it is not connected with the computational procedure since in all cases when the obtained data could be compared with experimental ones (dimers and oligomers [11]) a perfect fitting was obtained. The data generally agree as well with available computational results [10, 36, 38] so that we are facing some unknown peculiarity in the chemical behavior of the studied $sp^2$ nanocarbons. Obviously, the observed feature is connected with the specifics of intermolecular interaction between the components of the composites. In view of this, a question arises how correct is our suggestion concerning the superposition of deformational and chemical contributions into the total coupling energy. However, a standard view of $E_{def}^{tot}(R_{CC})$ and $E_{cov}^{tot}(R_{CC})$ plottings in figures 3, 13, and 20 appears to support this suggestion. If the suggestion is true, might be, a new third contribution, which so much effects $E_{cov}^{tot}$, should be taken into account Actually, as seen in figure 21, where the three $E_{cov}^{tot}(R_{CC})$ dependencies are joined together, there is a feeling that each of them well reproduces the others but is shifted along the energy scale. The considered nanocarbons are peculiar species, obviously not studied up to the bottom. This concerns, by example, topochemical character of addition reactions mentioned in section 3. Concerning the current case such a peculiarity of the species as their topology has not been so far taken into account. At the same time, one cannot exclude revealing some peculiar components of the intermolecular interaction similar to Casimir's repulsion in topological insulators [46] since the studied nanobuds are evidently formed under different topological conditions.

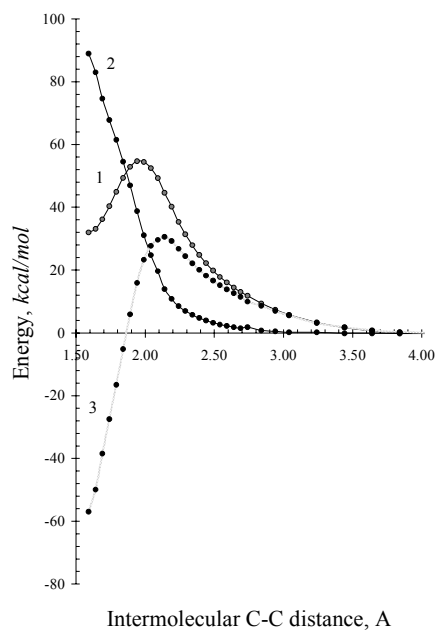 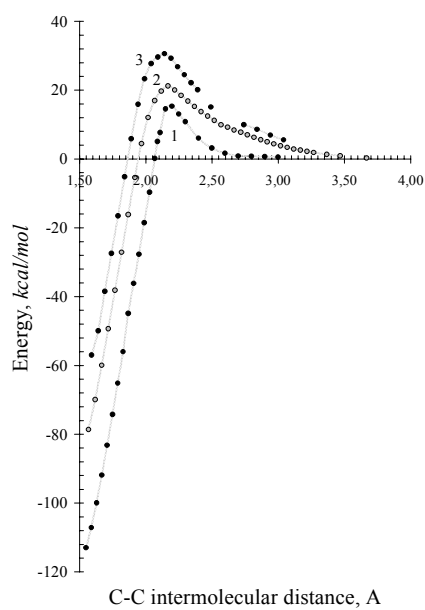

**Figure 20.** Profile of the barrier of the [$C_{60}$-(9,8)] GNB formation. 1. $E_{cpl}^{tot}$; 2. $E_{def}^{tot}$; 3. $E_{cov}^{tot}$.

**Figure 21.** $E_{cov}^{tot}$ plotting for $C_{60}$ dimer (1), [$C_{60}$-(4,4)] CNB 5 (2), and [$C_{60}$-(9,8)] GNB 8 (3).

**Table 4**. Joint characteristics of the nanobud [2+2] cycloadditions  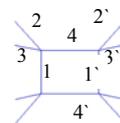

| Nanobuds | C-C bonds, Å[1] | | | | $N_{DA}^2$, e | $E_{chem}^{tot}$, kcal/mol | $E_{barr}^{NB}$, kcal/mol |
|---|---|---|---|---|---|---|---|
| | 1 | 2 | 3 | 4 | | | |
| $C_{60}+C_{60}$[3] | 1.548 | 1.515 | 1.515 | 1.596 | | | |
| | 1.548 | 1.516 | 1.516 | 1.596 | 0.271, 0.271 | -112.97 | 65.24 |
| $C_{60}+CBN^4$ | 1.567 | 1.483 | 1.486 | 1.652 | | | |
| 3 | 1.567 | 1.520 | 1.518 | 1.590 | 0.308, 0.307 | -134,31 | - |
| $C_{60}+CBN^4$ | 1.566 | 1.484 | 1.486 | 1.652 | | | |
| 5 | 1.566 | 1.520 | 1.518 | 1.590 | 0.305, 0.304 | -78,59 | 36.07 |
| $C_{60}+GNT^5$ | 1.591 | 1.496 | 1.496 | 1.581 | | | |
| 3 | 1.591 | 1.519 | 1.518 | 1.579 | 0.183, 0.173 | -98,91 | - |
| $C_{60}+GNT^5$ | 1.589 | 1.493 | 1.494 | 1.578 | | | |
| 6 | 1.589 | 1.519 | 1.517 | 1.580 | 0.216, 0.201 | -124,11 | - |
| $C_{60}+GNT^5$ | 1.590 | 1.492 | 1.492 | 1.578 | | | |
| 7 | 1.589 | 1.517 | 1.519 | 1.580 | 0.189, 0.165 | -60,42 | - |
| $C_{60}+GNT^6$ | 1.591 | 1.494 | 1.494 | 1.576 | | | |
| 8 | 1.592 | 1.519 | 1.518 | 1.580 | 0.227, 0.174 | -56.98 | 22.70 |

[1] The bond numeration corresponds to the insert. Two-row presentation distinguishes the primed bonds (the second rows) related to $C_{60}$ in all cases from unprimed ones (the first rows) related to the $C_{60}$ partner.
[2] The data are related to the pair of atoms of $C_{60}$ partner's in the nanobuds. The data for fullerene partner are presented in the first row.
[3] $C_{60}$ dimer.
[4] [$C_{60}$-(4,4)] CNB. CNB' numbering corresponds to that in figures 8 and 9
[5] [$C_{60}$-(5,5)] GNB. GNB' numbering corresponds to that in figures 16 and 19.
[6] [$C_{60}$-(9,8)] GNB (see figure 20).